# Monitoring the formation of nanowires by line-of-sight quadrupole mass spectrometry: a comprehensive description of the temporal evolution of GaN nanowire ensembles

Sergio Fernández-Garrido,* Johannes K. Zettler, Lutz Geelhaar, and Oliver Brandt

*Paul-Drude-Institut für Festkörperelektronik, Hausvogteiplatz 5–7, 10117 Berlin, Germany*

E-mail: garrido@pdi-berlin.de

## Abstract

We use line-of-sight quadrupole mass spectrometry to monitor the spontaneous formation of GaN nanowires on Si during molecular beam epitaxy. We find that the temporal evolution of nanowire ensembles is well described by a double logistic function. The analysis of the temporal evolution of nanowire ensembles prepared under a wide variety of growth conditions allows us to construct a growth diagram which can be used to predict the average delay time that precedes nanowire formation.

---

*To whom correspondence should be addressed

The growth of compound semiconductors in the form of nanowires (NWs) instead of films lifts the most fundamental constraint in epitaxial growth, namely, the necessity of using a substrate with a compatible crystal structure as well as similar in-plane lattice constants and thermal expansion coefficients. This advantage is due to the large aspect ratio of NWs. In contrast to epitaxial films, dislocations forming at the interface to the substrate do not propagate along the NW axis but remain at the interface or bend toward the free-sidewall surfaces.[1,2] A prominent example is the spontaneous formation of GaN NWs in plasma-assisted molecular beam epitaxy (PA-MBE), where dense arrays of single crystalline NWs form on crystalline as well as amorphous substrates.[3–7] Therefore, the growth of GaN in the form of NWs is envisioned as an ideal approach to integrate GaN-based electronic and opto-electronic devices with Si technology.[8–11]

In PA-MBE, GaN NWs form spontaneously under N-excess at elevated substrate temperatures without the need of using metal particles to collect the precursors and induce uniaxial growth.[3,5,6,12–16] Independent of the substrate, GaN NWs crystallize in the wurzite crystal structure, elongate along the (0001) axis, and exhibit $(10\bar{1}0)$ sidewall facets.[3,5,6,16,17] The high-vacuum environment characteristic of PA-MBE does not only facilitate the synthesis of high-purity material but also enables the use of powerful in situ characterization tools, such as reflection high-energy electron diffraction (RHEED) and line-of-sight quadrupole mass spectrometry (QMS). RHEED allows one to detect the onset for the spontaneous formation of GaN NWs[18–20] and provide information on NW in-plane as well as out-of-plane orientation distribution.[17,21] However, this technique cannot be used to quantify the amount of deposited material. The latter is as important as difficult to control due to the high desorption rate of Ga adatoms at elevated substrate temperatures.[22] QMS is thus an ideal complementary characterization technique because it enables the assessment of the deposition rate by measuring the desorbing Ga flux.[15,23–30]

The in situ assessment of the deposition rate by QMS paves the way for investigating the temporal evolution of GaN NW ensembles without interrupting the growth to quantify the amount of deposited material. Nevertheless, despite of this great advantage, QMS has been mainly used to monitor the desorbing Ga flux and not to investigate, in a systematic fashion, how the deposi-tion rate depends on the growth parameters.[15,28,29] In fact, the analysis of the deposition rate is hampered by the lack of comprehensive growth models. The existing models so far only apply to either the nucleation or the elongation of GaN NW ensembles.[15,19,20,31–36] Therefore, none of these models is capable of providing a quantitative and comprehensive description of the temporal evolution of GaN NW ensembles.

In this work, we use QMS to investigate the temporal evolution of the deposition rate during the growth of GaN NWs on Si(111) substrates in PA-MBE. The correlation of the QMS data with the morphology of the samples allows us to identify the three different growth stages reported in the literature, namely: (i) the incubation stage that precedes NW formation, (ii) the nucleation of GaN NWs, and (iii) the elongation of the GaN NWs along the (0001) axis.[15,16,18–20,31,37–40] We found that the deposition rate can be well described by a double logistic function. In order to systematically investigate how the spontaneous formation of GaN NWs depends on the growth parameters, we use this function to model the QMS transient of samples grown under different conditions. The results reveal that the spontaneous formation of GaN NWs at elevated temperatures is not limited by the axial growth rate during the elongation stage but by the long delay time that precedes NW formation. Finally, the dependence of the average delay time for NW formation on the growth parameters is summarized in a growth diagram. This diagram can be used as a guide to control the spontaneous formation of GaN NWs in PA-MBE as well as to envision novel growth approaches that might result in an improvement in the quality of GaN NWs.

The samples used for this study were prepared on 2 inch Si(111) substrates by PA-MBE using a radio-frequency $N_2$ plasma source for active N, and a solid-source effusion cell for Ga. Cross-sectional scanning electron microscopy (SEM) of thick GaN(0001) films grown under slightly N- and Ga-rich conditions on GaN/$Al_2O_3$(0001) substrates at low temperatures (680°C) was used to calibrate the impinging fluxes $\Phi_{Ga}$ and $\Phi_N$ in GaN-equivalent growth rate units of nm/min.[22] A growth rate of 1 nm/min is equivalent to $7.3 \times 10^{13}$ adatoms $cm^{-2}$ $s^{-1}$. The desorbing Ga flux during the experiments $\Phi_{des}$ was monitored in situ by QMS (see supporting information). The QMS response was also calibrated in GaN-equivalent growth rate units as explained in detail in

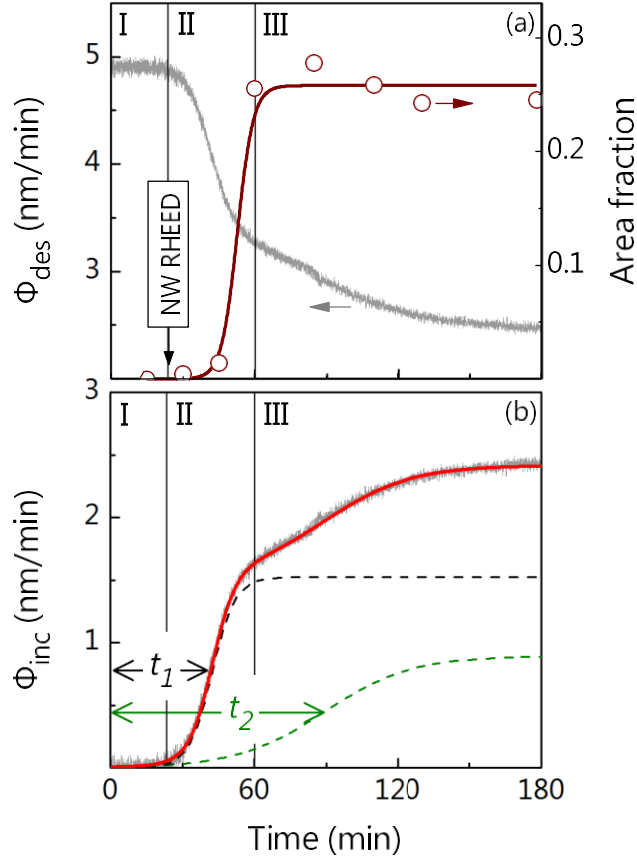

Figure 1: (a) Temporal evolution of both the desorbing Ga flux and the area fraction covered by GaN NWs during growth at 805°C with $\Phi_{Ga} = (4.9 \pm 0.6)$ nm/min and $\Phi_N = (10.5 \pm 0.5)$ nm/min. The solid maroon line is a guide to the eye. The time for the appearance of the first GaN-related spots in the RHEED pattern is also indicated in the figure. (b) Temporal evolution of the incorporation rate of Ga derived from the desorbing Ga flux as $\Phi_{Ga} - \Phi_{des}$. The solid red line displays the fit of the data by eq. 1. The dashed lines indicate the individual contribution of NW formation and collective effects to $\Phi_{inc}$. The average delay times for the onset of NW formation and collective effects are labeled as $t_1$ and $t_2$.

Refs. 24 and 25 as well as in the supporting information. Since there is no Ga accumulation on the substrate during the growth of GaN NWs,[15,16,37] the Ga incorporation rate per unit area $\Phi_{inc}$ (i. e., the deposition rate) was assessed as $\Phi_{Ga} - \Phi_{des}$. The growth temperature was measured with an optical pyrometer calibrated with the $1 \times 1 \rightarrow 7 \times 7$ surface reconstruction transition temperature of Si(111), namely, $\approx 860°C$.[41] The area fraction covered by GaN NWs was derived from the analysis of plan-view SEM images covering several hundreds of NWs by the open-source software ImageJ.[42] A detailed summary of all the samples investigated in this work can be found in the supporting information.

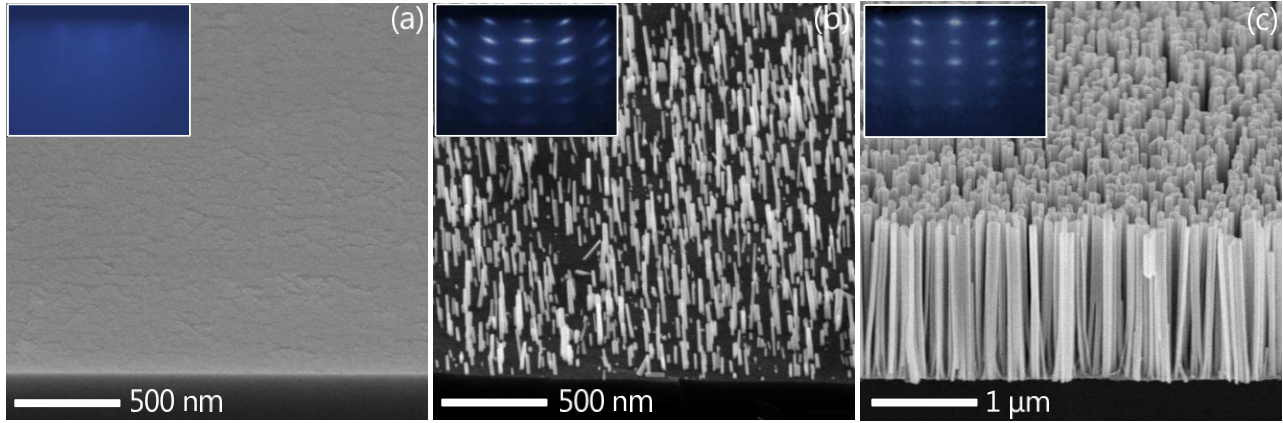

Figure 2: Bird's eye SEM images illustrating the morphology of the samples when the growth is interrupted at the three different stages depicted in Fig. 1. Figure (a) corresponds to the incubation stage, (b) to the nucleation stage, and (c) to the elongation stage. The insets show the corresponding RHEED patterns along the $[11\bar{2}0]$ azimuth.

Figures 1(a) and (b) show, respectively, the characteristic behavior of $\Phi_{des}$ and $\Phi_{inc}$ during the spontaneous formation and growth of GaN NWs at 805°C with $\Phi_{Ga} = (4.9 \pm 0.6)$ nm/min and $\Phi_N = (10.5 \pm 0.5)$ nm/min. In accordance with previous reports,[15,28,29] we distinguish the three different stages explained below and illustrated by the SEM images shown in Fig. 2.

During the first stage, the impinging Ga flux is fully desorbed. Consequently, $\Phi_{inc}$ is negligible within the experimental sensitivity of the QMS. As shown in the inset of Fig. 2(a), during this stage we observe neither GaN NWs in the SEM image nor GaN-related spots in the RHEED pattern. The latter is dim and diffuse due to the formation of amorphous $Si_xN_y$.[3,19,28,40] This stage is known in the literature as the incubation period, during which stable GaN nuclei have not yet been formed on the substrate.[18–20,28,31,38,39]

The second stage begins with the appearance of GaN-related spots in the RHEED pattern [see inset of Fig. 2(b)], which reveals the formation of stable GaN nuclei.[19,20] As reported before, these nuclei grow as spherical cap-shaped 3D islands until reaching a certain critical radius ($\approx$ 5nm).[19] Then, a shape transformation towards the final NW-like morphology occurs.[19] Afterwards, the NWs elongate along the (0001)-axis and grow radially until reaching a certain equilibrium radius that depends on the ratio between the impinging Ga and N fluxes.[15] At this point, the arc-shaped

GaN reflections observed in the RHEED pattern [see inset of Fig. 2(b)] evidence a broad out-of-plane orientation distribution. In the SEM image we observe well developed NWs, short ones, and many areas where NWs did not form yet. Therefore, each NW has its own incubation time. The continuous nucleation of GaN NWs leads to a broad initial height distribution[36] and to the rapid but progressive increase of $\Phi_{inc}$ [see Fig. 1(b)]. As depicted in Fig. 1(a), this second stage ends once the area fraction covered by GaN NWs saturates. At that point, the nucleation phase is completed.

The onset of the last stage is characterized by a clear change in the slope of the $\Phi_{inc}$ curve [see Fig. 1]. After this change, $\Phi_{inc}$ continuously increases until approaching a certain steady-state value. During this final stage, the arc-shaped reflections in the RHEED pattern evolve towards well-defined spots [see inset of Fig. 2(c)]. This change is due to the narrowing of the NW out-of-plane orientation distribution caused by the coalescence of closely spaced NWs.[43] The residual increase in $\Phi_{inc}$ observed throughout this stage is attributed to the time-dependent variation in the axial growth rate of individual NWs induced by collective effects, namely, the shadowing of the impinging fluxes by long NWs and the exchange of Ga atoms between adjacent NWs.[36] This latter effect is caused by the desorption and adsorption of Ga atoms at the NW sidewalls, as explained in detail in Ref. 36. Both collective phenomena tend to decrease $\Phi_{des}$ in comparison to the expected value for an isolated NW. For isolated or well separated NWs [see SEM image Fig. 2(b)] a significant amount of Ga atoms impinge directly on the substrate or on the NW sidewalls at distances to the NW top facet larger than their diffusion length (40–45 nm).[33,39,44] Consequently, most of these Ga atoms are desorbed and do not contribute to the growth. In contrast, for a dense NW ensemble, like the one shown in Fig. 2(c), most Ga atoms can contribute to the growth because they directly impinge or are re-adsorbed near the NW top facets. Consequently, $\Phi_{inc}$ must increase until the NW ensemble becomes homogeneous in height.[36]

We found that the temporal evolution of $\Phi_{inc}$ can be described by the following empirically

motivated equation:

$$\Phi_{\text{inc}} = \frac{A_1}{1 + \exp\left(-\frac{t - t_1}{\tau_1}\right)} + \frac{A_2}{1 + \exp\left(-\frac{t - t_2}{\tau_2}\right)}. \tag{1}$$

This equation is formed by the sum of two logistic functions and, as shown in Fig. 1(b), yields an excellent fit of the experimental data. The logistic function was originally proposed to model population growth and is commonly used to describe a wide variety of biological, geological, physical and chemical processes characterized by an initial exponential increase that eventually becomes limited due to the onset of competing effects.[45–48] In material science, this function has been recently proposed to describe the nucleation and growth of two-dimensional islands during molecular beam epitaxy.[48] In the present context, the first logistic function describes the NW formation stages, i.e., the incubation and shape transformation periods.[19,20] There, $A_1$ represents the deposition rate at the end of the nucleation stage, $t_1$ is the average delay time for NW formation, and $1/\tau_1$ a rate constant proportional to the NW formation rate after the formation of the first GaN NWs. Analogously, the second logistic function represents the temporal evolution of the collective effects as the NW ensemble becomes more homogeneous in height. Consequently, $A_2$ represents the further final increase in the deposition rate caused by collective effects, $t_2$ the average delay time for their onset, and $1/\tau_2$ a rate constant that reflects the temporal variation in the contribution of collective effects to the total deposition rate. Therefore, in contrast to previous growth models,[15,19,20,31–36] eq. 1 provides an empirical but comprehensive description of the entire growth process (i.e., from the beginning to the end of the growth).

Having identified eq. 1 as a suitable expression to describe the temporal evolution of $\Phi_{\text{inc}}$, we can now investigate in a systematic fashion how the spontaneous formation and growth of GaN NWs depends on the growth parameters, namely, the substrate temperature, $\Phi_{\text{Ga}}$, and $\Phi_{\text{N}}$.

Figure 3 shows the temperature dependence of $t_1$ for a series of samples grown with $\Phi_{\text{Ga}} = (4.9 \pm 0.6)$ nm/min and $\Phi_{\text{N}} = (10.5 \pm 0.5)$ nm/min. As can be seen, $t_1$ increases exponentially with the substrate temperature by more than one order of magnitude between 775 and 835°C. The

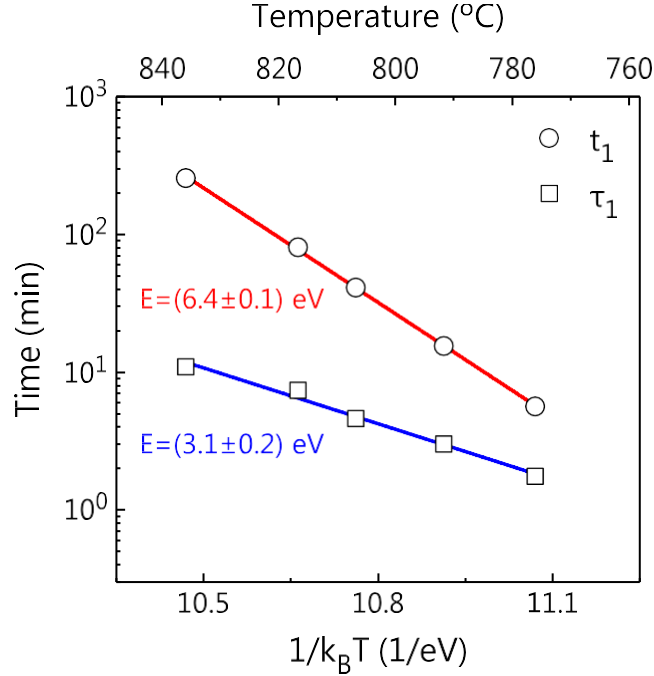

Figure 3: Temperature dependencies of the delay time for NW formation $t_1$ and the time constant $\tau_1$. All samples were grown with $\Phi_{Ga} = (4.9 \pm 0.6)$ nm/min and $\Phi_N = (10.5 \pm 0.5)$ nm/min. The solid lines are fits of the data by an Arrhenius law.

temperature dependence of $t_1$ can be properly described by an Arrhenius law:

$$t_1 = C_1 \exp(-E_N/k_B T), \qquad (2)$$

where $C_1$ is a constant. The best fit of eq. 2 to the experimental data yields an activation energy $E_N$ of $(6.4 \pm 0.1)$ eV. This value is significantly higher than that of $(4.9 \pm 0.1)$ eV determined in Ref. 20 using RHEED. However, while $t_1$ represents the *average* delay time for NW formation, the incubation time reported in Ref. 20 represents the delay time for the formation of the *first* GaN NWs. Note that, as indicated in Fig. 1, these two times can be quite different. Therefore, even though in both cases the activation energy is supposed to be related to the GaN nucleation barrier,[20] the actual values cannot be directly compared.

Figure 3 also presents the variation of $\tau_1$ with the substrate temperature for the series of samples introduced before. Similarly to $t_1$, this parameter increases exponentially with substrate temperature, reflecting a decrease in the NW formation rate. However, $\tau_1$ is always much lower than $t_1$.

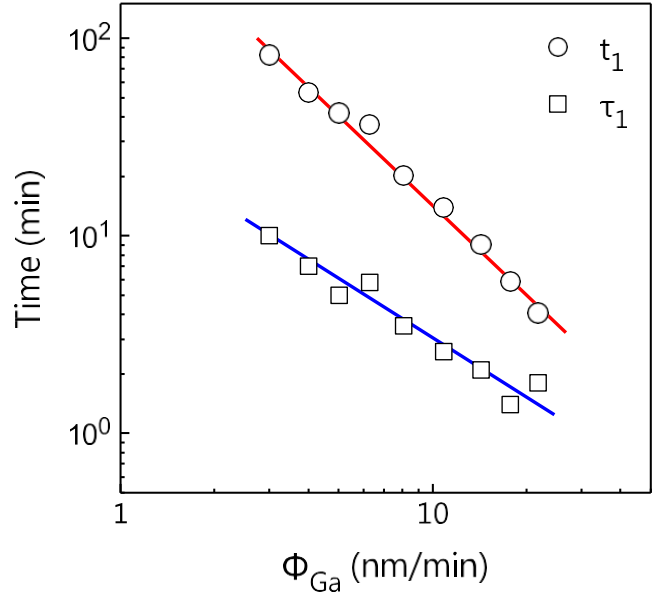

Figure 4: Variation with $\Phi_{Ga}$ of the delay time for NW formation $t_1$ and the time constant $\tau_1$. All samples were grown at 805°C with $\Phi_N = (10.5 \pm 0.5)$ nm/min. The solid lines are fits of the data by eqs. 3 and 4.

The temperature dependence of $\tau_1$ can also be described by an Arrhenius law and the activation energy derived from the fit is $(3.1 \pm 0.2)$ eV. This value is remarkably lower than the one estimated in Ref. 29 for Ga desorption during the growth GaN NWs, namely, $(4.0 \pm 0.3)$ eV.

To examine the impact of the impinging Ga flux on the formation of GaN NWs, we analyzed a second series of samples grown using different $\Phi_{Ga}$ values (3–21.7 nm/min) at 805 °C with $\Phi_N = (10.5 \pm 0.5)$ nm/min. Figure 4 shows the variation of $t_1$ with $\Phi_{Ga}$. The correlation is clear: the lower the Ga flux, the longer the average delay time for NW formation. As shown in the figure, the variation of $t_1$ with $\Phi_{Ga}$ follows the expression:

$$t_1 = \frac{C_2}{\Phi_{Ga}^{3/2}} \qquad (3)$$

where $C_2$ is a constant. This dependence is consistent with the nucleation studies reported in Ref. 20, in which an identical power dependence was reported for the time required to observe the first GaN-related spots in the RHEED pattern. Such a power law was explained within the framework of the standard island nucleation theory and the 3/2 exponent was related to the critical size of the

stable GaN nuclei.[20]

Figure 4 also shows the variation of $\tau_1$ with the impinging Ga flux. As in the case of $t_1$, this parameter steadily decreases when the Ga flux is increased. The plot evidences a clear power law dependence given by:

$$\tau_1 = \frac{C_3}{\Phi_{Ga}}, \quad (4)$$

where $C_3$ is a constant. These results demonstrate that a higher Ga flux does not only decrease the average delay time for NW formation but also increases the formation rate after the nucleation of the first GaN NWs.

The assessment of the average delay time required for the formation of the GaN NWs $t_1$ allows us to estimate the average axial growth rate during the elongation stage. This quantity can be calculated as the average NW length at the end of the growth divided by the actual elongation time, i.e., the total growth time minus the average delay time for NW formation.

Figure 5(a) shows the dependence of the average axial growth rate on the substrate temperature. Clearly, the average axial growth rate is always higher than $\Phi_{Ga}$ and steadily increases with the substrate temperature until approaching the value of $\Phi_N$. In agreement with previous reports,[15,21,31,33,39,44,49] these results evidence a significant diffusion of Ga adatoms along the side facets toward the NW tips. In contrast, N diffusion seems to be almost negligible, as already pointed out in Ref. 15. According to the quantitative growth model presented in Ref. 33, the increase in the axial growth rate with increasing substrate temperature is due to the enhanced diffusion length of Ga adatoms along the NW sidewalls. Interestingly, unlike in this previous work, we do not observe a drop in the average axial growth rate above 800°C. This discrepancy can be understood by taking into account that the axial growth rates reported in Ref. 33 were assessed considering not the average delay time for NW formation but the delay time for the appearance of the GaN-related spots in the RHEED pattern. Obviously, according to the results shown above, such an approach tends to underestimate the average axial growth rate, especially at elevated substrate temperatures due to the larger values of $\tau_1$.

The dependence of the average axial growth rate on the impinging Ga flux at 805°C is dis-played in Fig. 5(b). As can be seen in the figure, the average axial growth rate is basically independent of the impinging Ga flux. These results evidence that, despite the N excess required for the spontaneous formation of GaN NWs in PA-MBE,[15] at elevated substrate temperatures the axial growth rate may become N-limited. These results agree well with the experiments reported in Refs. 15,21,49–53.

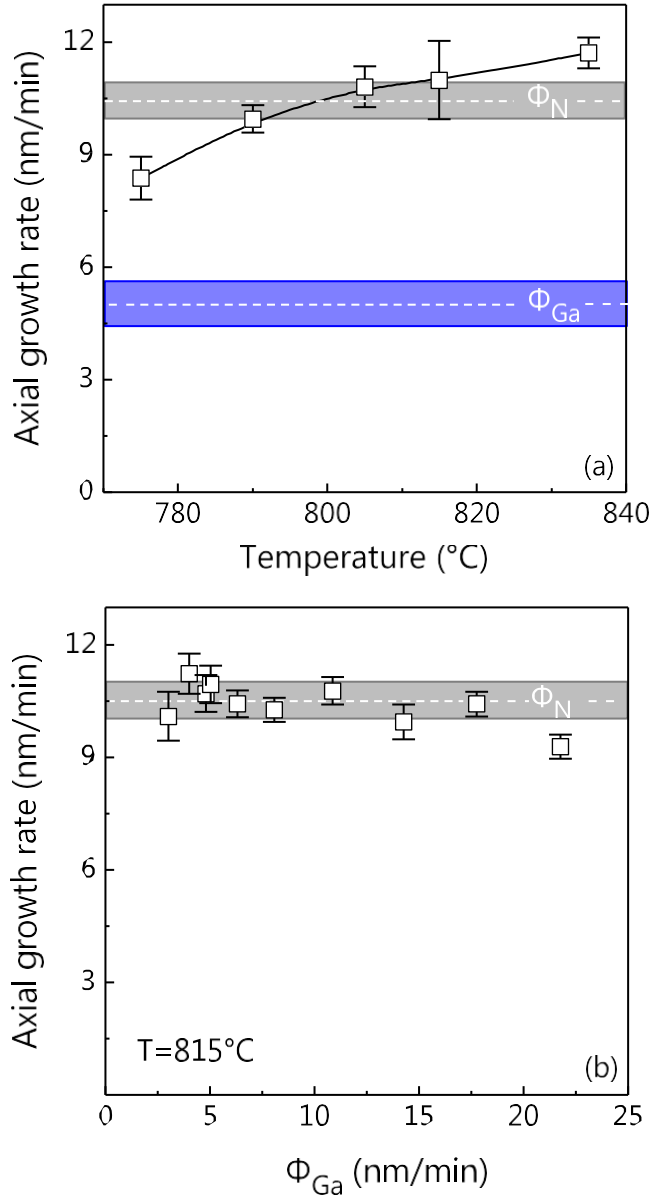

Figure 5: (a) Temperature dependence of the average axial growth rate of GaN NWs grown with $\Phi_{Ga} = (4.9 \pm 0.6)$ nm/min and $\Phi_N = (10.5 \pm 0.5)$ nm/min. The solid line is a guide to the eye. (b) Variation in the average axial growth rate of GaN NWs grown at 805°C with $\Phi_N = (10.5 \pm 0.5)$ nm/min. In the figures the dashed lines indicate the impinging N and Ga fluxes.

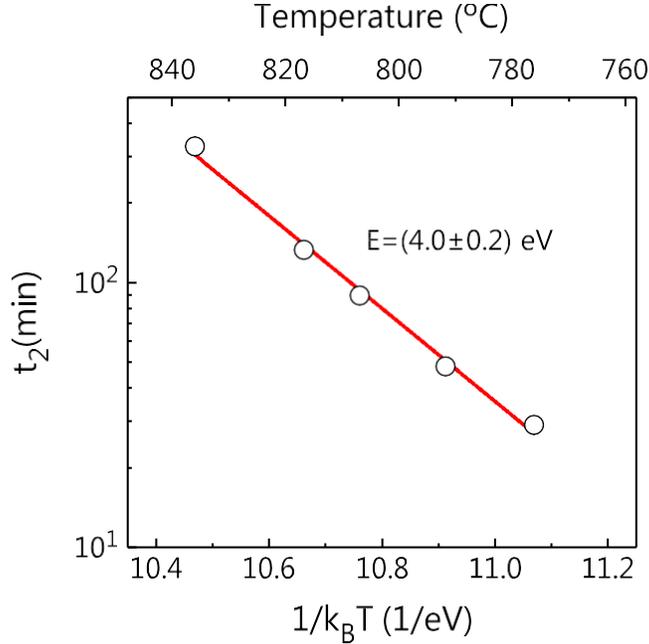

Figure 6: Temperature dependence of the average delay time for the onset of collective effects $t_2$. The solid line is a fit of the data by an Arrhenius law. All samples were grown with $\Phi_{Ga} = (4.9 \pm 0.6)$ nm/min and $\Phi_N = (10.5 \pm 0.5)$ nm/min.

For the two previous series of samples, we also analyzed how the parameters related to the collective effects in eq. 1, namely $t_2$ and $\tau_2$, depend on the substrate temperature and the impinging Ga flux. As shown in Fig. 6, the substrate temperature has a strong impact on the average delay time for the onset of collective effects, i. e., $t_2$. This parameter increases exponentially with substrate temperature by approximately one order of magnitude between 775°C and 835°C. The temperature dependence is properly described by an Arrhenius law with an activation energy of $(4.0 \pm 0.2)$ eV. This value is thus comparable to the activation energy reported in Ref. 24 for the desorption of Ga adatoms from a $(000\bar{1})$GaN surface (3.7 eV). The parameter $\tau_2$ was found to increase within the investigated temperature range from 12 to 34 min but the experimental data do not exhibit a clear

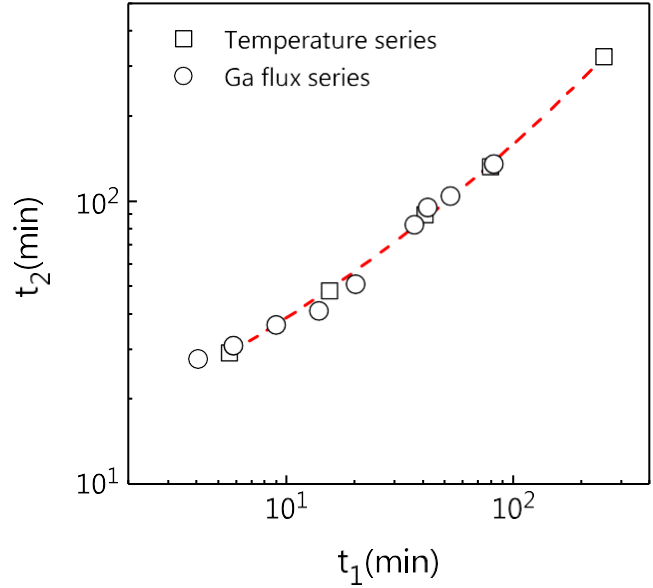

Figure 7: Correlation between the average delay times for NW formation $t_1$ and the onset of collective effects $t_2$. The open squares correspond to a series of samples grown at different temperatures with $\Phi_{Ga} = (4.9 \pm 0.6)$ nm/min. The open circles correspond to a series of samples grown at 805°C using different values of $\Phi_{Ga}$. For both series of samples, $\Phi_N = (10.5 \pm 0.5)$ nm/min. The dashed line is a guide to the eye.

Arrhenius like temperature dependence (see supporting information). Regarding the dependence of $t_2$ and $\tau_2$ on the impinging Ga flux, we observed that the higher the Ga flux the shorter the times (see supporting information). Specifically, $t_2$ decreases from 136 to 22 min and $\tau_2$ from 36 to 12 min when the Ga flux is increased from 3 to 21.7 nm/min. As shown in the supporting information, similarly to $t_1$ and $\tau_1$, the dependencies of $t_2$ and $\tau_2$ on the impinging Ga flux can be described by power laws but with different exponents as well as larger error margins. Nevertheless, the present results evidence that the time required to form a homogeneous NW ensemble after the nucleation stage depends on the substrate temperature as well as on the impinging Ga flux.

The comparison of the average delay times for NW formation $t_1$ and for the onset of collective effects $t_2$ evidences a clear correlation. As shown in Fig. 7, where $t_2$ is plot as a function of $t_1$ for the two series of samples discussed above, $t_2$ monotonically increases with $t_1$. This is indeed the expected trend according to the interpretation of the two logistic functions that comprise eq. 1. Note that $t_2$ scales with $t_1$ because as far as the NW number density remains below a certain threshold value, the exchange of Ga atoms between adjacent NWs as well as the shadowing of the impinging fluxes by long NWs are negligible processes. Consequently, the shorter the average delay time for NW formation, the earlier the onset of collective effects.

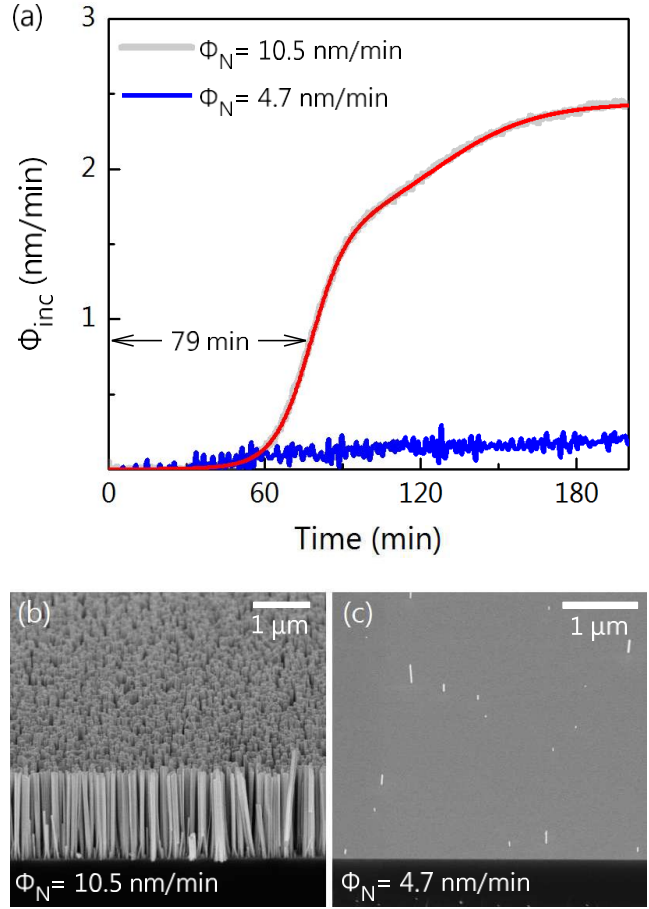

Figure 8: (a) Temporal evolution of the incorporation rate of Ga for two samples grown at 815°C with $\Phi_{Ga} = (4.9 \pm 0.6)$ nm/min using different values of $\Phi_N$, namely, $(10.5 \pm 0.5)$ and $(4.9 \pm 0.5)$ nm/min. The solid red line represent a fit to eq. 1. The average delay time for NW formation $t_1$ derived from the fit is 79 min. (b) and (c) are bird-eyes SEM micrographs of the samples grown with $\Phi_N = (10.5 \pm 0.5)$ and $(4.7 \pm 0.5)$ nm/min, respectively.

To complete our study, we have also investigated whether the impinging active N flux plays a role in the spontaneous formation of GaN NWs. Figure 8 shows the temporal evolution of the Ga incorporation rate as well as the final morphology of two samples grown using quite different active N fluxes, namely, $(10.5 \pm 0.5)$ and $(4.7 \pm 0.5)$ nm/min. For both samples, the substrate temperature, the impinging Ga flux, and the growth time were 815°C, $(4.9 \pm 0.6)$ nm/min, and 200 min, respectively. For the high active N flux, the temporal evolution of $\Phi_{inc}$ exhibits the three different growth stages described above. However, for the low active N flux, $\Phi_{\text{inc}}$ remains almost constant during growth. The SEM image shown in Fig. 8(b) reveals a dense and homogeneous NW ensemble for the sample grown with $\Phi_{\text{N}} = (10.5 \pm 0.5)$ nm/min. In contrast, for the sample grown with $\Phi_{\text{N}} = (4.7 \pm 0.5)$ nm/min, there are only a few sparse NWs [see Fig. 8(c)]. Therefore, as expected from the temporal evolution of $\Phi_{\text{inc}}$, the nucleation stage was not finished for the sample grown with the low active N flux. Consequently, we conclude that the active N flux does not only influence the axial growth rate but also plays an important role during the nucleation stage, where a higher active N flux results in a significant reduction in the average delay time for NW formation.

As further discussed below, the results presented throughout this manuscript allow us to establish the limits for the spontaneous formation of GaN NWs and to envision novel growth approaches that might result in an improvement in the quality of GaN NWs.

In general, in crystal growth the concentration of point defects as well as the incorporation of impurities decrease with increasing substrate temperature. In the case of the growth of GaN films in PA-MBE, the maximum growth temperature that can be achieved is limited by GaN decomposition.[27,54] Since under high vacuum the GaN decomposition rate becomes comparable to the available impinging fluxes for temperatures above 750°C,[27] the typical growth temperatures reported in the literature are well below 800°C.[22,26,55] This temperature is more than $200 - 300$°C lower than the optimal growth temperatures used in other epitaxial growth techniques, such as hydride vapor phase epitaxy (HVPE) or metalorganic chemical vapor deposition (MOCVD).[56–58] The study presented here allows us to conclude that, in contrast to films, GaN decomposition does not limit the elongation rate of GaN NWs within the investigated temperature range ($775 - 835$°C). In fact, once NWs are formed, they elongate quite rapidly with a rate that is close to the impinging active N flux. This might be partly due to the efficient reduction of the effective GaN decomposition rate caused by the N excess.[27] Therefore, we conclude that the maximum growth temperature that can be achieved is not limited by the elongation stage but by the long delay time that precedes the spontaneous formation of GaN NWs.

As shown above, the average delay time for the spontaneous formation of GaN NWs during the nucleation stage depends on the substrate temperature as well as on the impinging fluxes, $\Phi_{Ga}$ and $\Phi_N$. By combining eqs. 2 and 3, the average delay time for NW formation can be written in a compact fashion as a function of the substrate temperature and the impinging Ga flux:

$$t_1 = C_4 \frac{\exp(-E_N/k_B T)}{\Phi_{Ga}^{3/2}}, \tag{5}$$

where $C_4$ is a constant that depends on the impinging active N flux.

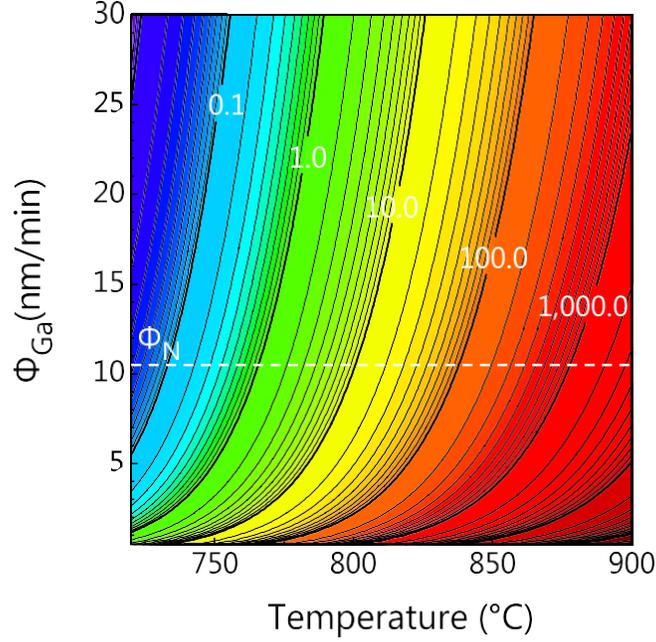

Figure 9: Growth diagram depicting the dependence of the average delay time for NW formation $t_1$ (in minutes) on the growth temperature and the impinging Ga flux for a given active N flux of $(10.5 \pm 0.5)$nm/min. The latter is indicated by the horizontal dashed line. The time $t_1$ is displayed as a contour plot with a logarithmic scale.

Figure 9 depicts a growth diagram, derived from eq. 5, which visualizes the impact of the substrate temperature as well as the impinging Ga flux on the average delay time for NW formation. In the diagram, the value of $t_1$ is displayed as a contour plot with a logarithmic scale. Despite having investigated only Ga fluxes lower than 22 nm/min and substrate temperatures between 775 and 835 °C, eq. 5 allows us to predict the values of $t_1$ for unexplored growth conditions. As can be clearly seen, $t_1$ increases with increasing growth temperature but decreases with the impinging Ga flux. For very low values of $t_1$, NW formation takes place very fast. However, the combination of high impinging Ga flux and low substrate temperature typically results in either highly-coalesced NW ensembles or compact layers.[12] For high substrate temperatures or very low Ga fluxes, $t_1$ rapidly becomes as long as several tens of hours. Consequently, under these latter conditions, the growth of GaN NWs is impracticable. We stress that the present growth diagram is quantitatively valid only for an active N flux of 10.5 nm/min. As discussed above, higher active N fluxes can be used to reduce $t_1$ and would shift the entire growth diagram toward lower Ga fluxes and higher substrate temperatures.

The growth diagram shown in Fig. 9 implies that a simple way to grow GaN NWs at significantly higher temperatures than those previously reported in the literature[5,6,12,13] consists in leaving the commonly used growth regime of nominally N-rich growth conditions, i. e., $\Phi_{Ga}/\Phi_N < 1$. Instead, the use of nominally Ga-rich growth conditions enables the growth of GaN NWs at higher temperatures while maintaining a comparable morphology. Note that the NW-like morphology can be preserved because the growth may still take place under effective N excess, as required for the spontaneous formation of GaN NWs,[15] due to the exponential increase of the desorbing Ga flux with increasing substrate temperature.[22] We can also use eq. 5 to predict the maximum substrate temperature that can be achieved using this growth approach. For this estimate, we assume that $t_1$ depends on $\Phi_N$ in the same way as on $\Phi_{Ga}$, namely, $t_1 \propto 1/\Phi_N^{3/2}$. As obvious from eq. 5, the maximum substrate temperature is limited by the available impinging fluxes. In PA-MBE, the fluxes are limited by the necessity of maintaining a pressure low enough to guarantee a molecular beam regime (typically below $10^{-4}$ Torr) as well as by technical limitations, such as the efficiency of radio-frequency $N_2$ plasma sources. The highest active N fluxes reported so far in the literature are of the order of 43 nm/min[59] while the typical Ga fluxes provided by solid-source effusion cells are also of the order of several tens of nm/min. If we consider that delay times for NW formation longer than 600 min are impracticable, for impinging Ga and active N fluxes of the order of 50 nm/min the maximum achievable substrate temperature is about 960°C.

An alternative approach to achieve elevated substrate temperatures, that can also be combined with the use of nominally Ga-rich growth conditions, would consist in using growth schemes that favor NW nucleation. Enhancing nucleation can be achieved by either nucleating the NWs at a lower temperature than the one used during the elongation stage or inserting a buffer layer of a material that lowers the nucleation barrier, such as AlN[60,61] or amorphous Al$_x$O$_y$.[7] The combination of this approach with the use of nominally Ga-rich growth conditions is expected to allow one to synthesize GaN NWs at temperatures approaching 1000°C, comparable to values used in HVPE and MOCVD.

To summarize, the assessment of the deposition rate by measuring the desorbing Ga flux using QMS allowed us to monitor in situ the spontaneous formation of GaN NWs on Si(111) substrates during PA-MBE. We found that the temporal evolution of GaN NW ensembles can be phenomenologically described by a double logistic function. This simple equation provides a quantitative and comprehensive description of the entire growth process. The analysis of the deposition rate for a wide variety of growth conditions enabled us to determine the impact of the growth parameters on the spontaneous formation of GaN NWs. The results were summarized in a growth diagram that can be used as a guide to control the growth and properties of GaN NWs as well as to envision novel growth approaches.

## Associated content

Supporting information. Experimental additional details on QMS; summary of the samples used in this work as well as of the parameters derived from fitting eq. 1 to the QMS transients; temperature dependence of the time constant $\tau_2$; impact of the impinging Ga flux on the time constants $t_2$ and $\tau_2$. This material is available free of charge via Internet at http://pubs.acs.org.

## Acknowledgement

We would like to thank Anne-Kathrin Bluhm for providing the scanning electron micrographs presented in this work, Hans-Peter Schönherr for his dedicated maintenance of the MBE system, and Vladimir Kaganer for fruitful discussions and a critical reading of the manuscript. We are also indebted to Henning Riechert for continuous encouragement and support. Partial financial support of this work by the Deutsche Forschungsgemeinschaft within SFB 951 is gratefully acknowledged.

**Table of Contents Graphic**

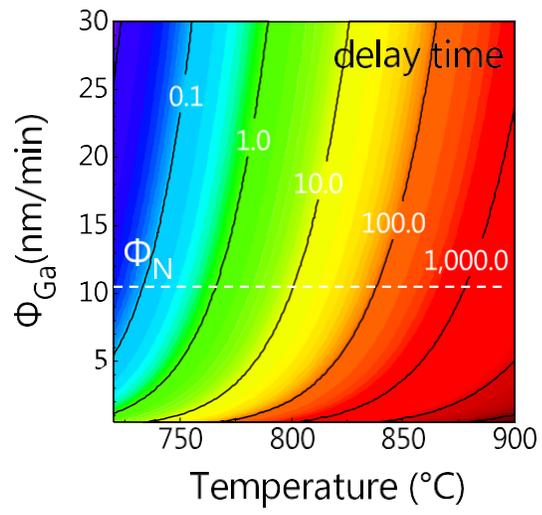